\documentstyle[12pt]{article}

\title{Perturbative Dynamics of Quantum  General Relativity }

\author{John Donoghue\\ [5mm]
Department of Physics and Astronomy,\\ University of
Massachusetts, 
Amherst, MA 01003 U.S.A.}
\date{}
\begin{document}
\begin{titlepage}
\maketitle
\begin{abstract}
The quantum theory of General Relativity at low energy 
exists and is of the form called "effective field theory".
In this talk I describe the ideas of effective field theory
and its application to General Relativity.
\end{abstract}
{\vfill Invited plenary talk at the Eighth Marcel Grossmann Conference on
General Relativity, Jerusalem, June 1997, to be published in the
proceedings.} 
\end{titlepage}

\section{Introduction}
The conference organizers originally suggested that the title of this
talk be: ``Gravity and the Quantum: The view from particle physics".
While it is presumptuous for me to claim to speak for all particle
physicists, there is in fact a widely held ``view'' within the
particle community that carries an important insight not always appreciated
within the gravity community. In this visual analogy, we see clearly 
a variety
of beautiful low-lying hills representing the Standard Model and its
applications in known physics. However, there are two sets of clouds
on the horizon which ultimately obscure our view. One cloud is located 
at 1 TeV, just beyond the reach of present accelerators.  Beyond this scale,
we expect to find the physics which governs electroweak symmetry 
breaking, with the expectation being that we will uncover new particles
and new interactions that change the way that we think of Nature. The
other cloud is more ominous, sitting at the Planck scale. Beyond this, we 
don't know what to expect, but it likely will be something totally new. 
Neither the Standard Model nor General Relativity is likely to emerge 
unchanged beyond this scale. So this ``view" recognizes that the 
scenery that we see (our low energy theory) 
is likely to change if we are ever able
to see beyond the ``clouds".

The insight behind this view suggests a new way of looking at
quantum General Relativity.  Since we only know that General Relativity
is valid at low energy, the key requirement is that the quantum theory 
can be applied to gravity {\it at present scales}.  What goes on beyond the
Planck scale is a matter of speculation, but gravity and quantum mechanics 
had better go together at the scales where they are both valid.

The good news is that the quantum theory of General Relativity at low energies
exists and is well behaved. It is of the form of a type of field theory called 
Effective Field Theory~\cite{eft}. 
This is true no matter what the ultimate high-energy 
theory turns out to be.  Given all the work that has gone into quantum gravity,
I feel that this is a significant result. 
The development of effective field theory is an important
part of the past decade, and anyone who cares about field theory should
learn about it. It has become a standard way of calculating within particle
physics, and the way of thinking is widely internalized in the younger
generation. In fact, it is would be a reasonably common expectation of young 
theorists that it is possible that
the gravitational effective field theory may turn out to be a better quantum
theory than the Standard Model, as the former 
may extend in validity all the way up to 
the Planck scale, while the Standard Model will likely be
fundamentally modified at 1 TeV.

This talk describes some of the features of the effective field theory of 
General Relativity~\cite{jfd1,jfd2}. 
The effective field theory completes a program for 
quantizing General Relativity that goes back to Feynman and 
De Witt~\cite{fdw}, and 
which has received contributions from many researchers over the years.
Earlier work focused on quantization and on the divergence structure at
high energy. The contribution of effective field theory is to shift the
focus back to the low energy where the theory is valid, and to classify 
the reliable predictions. The low energy quantum effects are distinct from 
whatever goes on at high energy. 
Of course, the effective theory does not answer
all the interesting questions that we have about the ultimate theory. 
However, in principle it answers all those questions that we have a right 
to know with our present state of knowledge about the content of the theory.
I will attempt to be clear about the limits of the effective theory as well as 
its virtues.

The outcome of this is that we need to stop spreading the falsehood that 
General Relativity and Quantum Mechanics are incompatible. They
go together quite nicely at ordinary energies. 
Rather, a more correct statement is that we do
not yet know the ultimate high energy theory in Nature. This change
in view is important for the gravity community to recognize, because it
carries the implication that the ultimate theory is likely to be something 
new, not just a blind continuation of General Relativity beyond the 
Planck scale

\section{Effective Field Theory}

First let's describe effective field theory in general. Once you
understand the basic ideas it is easy to see how it applies to
gravity. The phrase ``effective'' carries the connotation of a low
energy approximation of a more complete high energy theory. However,
the techniques to be described don't rely at all on the high energy
theory.  It is perhaps better to focus on a second
meaning of ``effective'', ``effective'' $\sim$ ``useful'', which
implies that it is
the most effective thing to do. This is because the particles and
interactions of the effective theory are the useful ones at that
energy. An ``effective Lagrangian'' is a local Lagrangian which
describes the low energy interactions.  ``Effective
field theory'' is more than just the use of effective Lagangians. It
implies a specific full field-theoretic treatment, with loops,
renormalization etc. The goal is to extract the full quantum effects
of the particles and interactions at low energies.

The key to the separation of high energy from low is the uncertainty
principle. When one is working with external particles at low energy,
the effects of virtual heavy particles or high energy intermediate
states involve short distances, and hence can be represented by a
series of local Lagrangians. A well known example is the Fermi 
theory of the weak interactions, which is a local effective Lagrangian
describing the effect of the exchange of a heavy W boson. 
This locality is true even for the high energy
portion of loop diagrams. An example of the latter is the high energy 
portion of the fermion self energy, which is equivalent to a mass 
counterterm in a local Lagrangian. In contrast, effects that are non-local,
where the particles propagate long distances, can only come from the
low energy part of the theory. The exchange of a massless 
photon at low energy can never be represented by a local
Lagrangian. From this distinction, we know that we
can represent the effects of the high energy theory by the {\it most
general} local effective Lagrangian. 

The second key is the energy
expansion, which orders the infinite number of terms within this most
general Lagrangian in powers of the low energy scale divided
by the high energy scale. To any given order in this small parameter,
one needs to deal with only a finite number of terms (with
coefficients which in general need to be determined from experiment).
The lowest order Lagrangian can be used to determine the propagators
and low energy vertices, and the rest can be treated as perturbations.
When  this theory is quantized and used to calculate loops, the usual
ultraviolet divergences will share the form of the most general
Lagrangian (since they are high energy and hence local) and can be
absorbed into the definition of renormalized couplings. There are
however leftover effects in the amplitudes from long distance
propagation which are distinct from the local Lagrangian and which are
the quantum predictions of the low energy theory.

This technique can be used in both renormalizable and
non-renormalizable theories, as there is no need to restrict the
dimensionality of terms in the Lagrangian. (Note that the terminology
is bad: we {\it are} able to renormalize non-renormalizable theories!) 
Renormalizable theories are a particularly compact and predictive 
class of theories. However, many physical  effects require 
non-renormalizable interactions and these need not 
destroy the quantum theory. In fact, a common calculational 
device is to isolate the relevant interactions only, even if 
this implies a non-renormalizable theory, and to use the techniques 
of effective field theory to perform a simpler calculation 
than if one were to compute using the full theory.
This is done in Heavy Quark Effective Theory~\cite{hqet} as well
as in the theory of electroweak radiative corrections. 

The
effective field theory which is most similar to general relativity is
chiral perturbation theory~\cite{cpth}, 
which describes the theory of pions and
photons which is the low energy limit of QCD. The theory is highly
nonlinear, with a lowest order Lagrangian which can be written with
the exponential of the pion fields
\begin{eqnarray}
{\cal L} & = & {F_\pi^2 \over 4} Tr\left( \nabla_\mu U \nabla^\mu
U^\dagger \right) + {F_\pi^2 m_\pi^2 \over 4} Tr\left( U+U^\dagger
\right) \nonumber \\
U & = & exp \left(i {\tau^i \pi^i(x) \over F_\pi} \right) \ \ ,
\end{eqnarray}
\noindent with $\tau^i$ being the SU(2) Pauli matrices and $F_\pi
=92.3 MeV$ being a dimensionful coupling constant. This shares with
general relativity the dimensionful coupling, the non-renormalizable
nature and the intrinsically nonlinear Lagrangian. This theory has
been extensively studied theoretically, to one and two loops, and
experimentally. There are processes which clearly reveal the
presence of loop diagrams. In my talk, I displayed some of the 
predictions and experimental tests
of chiral perturbation theory, most of which were taken
from the book published with my co-authors~\cite{eft}. The point was to
illustrate the fact that effective field theory is not just an idea,
but is a practical tool that is applied in real-world physics. 
In a way, chiral perturbation theory is the
model for 
a complete non-renormalizable effective field theory in the same way
that QED serves as a model for renormalizable field theories.

\section{Overview of the gravitational effective field theory}

At low energies, general relativity automatically behaves in the way
that we treat effective field theories. This is not a philosophical
statement implying that there must be a deeper high energy theory of which
general relativity is the low energy approximation. 
Rather, it is a practical statement. Whether or not 
general relativity is truly fundamental, the low energy quantum
interactions must behave in a particular way because of the nature of
the gravitational couplings, and this way is that of effective field
theory. 

The Einstein action, the scalar curvature, involves two
derivatives on the metric field. Higher powers of the curvature,
allowed by general covariance, involve more derivatives and hence the
energy expansion has the form of a derivative expansion. 
The higher powers of the curvature in the most general Lagrangian do
not cause problems when treated as low energy perturbations~\cite{simon}.
The Einstein action is in fact readily quantized, using gauge-fixing
and ghost fields ala Feynman, DeWitt, Faddeev, Popov~\cite{fdw}. 
The background
field method used by 'tHooft and Veltman~\cite{thv} is most beautiful in this
context because it allows one to retain the symmetries of general
relativity in the background field, while still gauge-fixing the
quantum fluctuations.  The applications of these methods allow 
the quantization of general relativity in as straight-forward a 
way as QCD is quantized.  

The problem with the field theory program comes not at the 
level of quantization, but in attempting to make meaningful
calculations.
The dimensionful nature of the gravitational
coupling implies that loop diagrams (both the finite and infinite
parts) will generate effects at higher orders in the energy
expansion~\cite{dt1}. 
In previous times when we only understood renormalizable 
field theory, this was a problem because the divergences could
not be dealt with by a renormalization of the original Lagrangian.
However, in effective field theory, one allows a more 
general Lagrangian. Since the divergences come from the 
high energy portion of loop integrals, they will be equivalent 
to a local term in a Lagrangian. Since the effective Lagrangian 
allows all terms consistent with the theory, 
and each term is governed by one or more
parameters describing its strength, there is a parameter available
corresponding to each divergence. We absorb the high energy
effects of  the loop diagram into a renormalized parameter, which also
contains other unknown effects from the ultimate high energy theory.
The one and two loop counterterms for graviton
loops are known~\cite{thv,twoloop} and, as expected, go into 
the renormalization of the coefficients in
the Lagrangian. However, these are not really predictions of the
effective theory. The real action comes at low energy.

How in practice does one separate high energy from low? Fortunately,
the calculation takes care of this automatically, although it is
important to know what is happening. Again, the main point is that the
high energy effects share the structure of the local Lagrangian, while
low energy effects are different. When one completes a calculation,
high energy effects will appear in the answer in the same way that the
coefficients from the local Lagrangian will. One cannot
distinguish these effects from the unknown coefficients. However, low
energy effects are anything that has a different structure. Most
often the distinction is that of analytic versus non-analytic in
momentum space. Analytic expressions can be Taylor expanded in the
momentum and therefore have the behavior of an energy expansion, much
like the effects of a local Lagrangian ordered in a derivative
expansion. However, non-analytic terms can never be confused with the
local Lagrangian, and are intrinsically non-local. Typical
non-analytic forms are $\sqrt{-q^2}$ and $\ln(-q^2)$. These are always
consequences of low energy propagation. 

Having provided this brief overview of the way that effective field
theory may be applied to general relativity, let me be a bit more
explicit about some of these steps.

\section{The energy expansion in general relativity}

What is
the rationale for choosing the gravitational action proportional to
$R$ and
only $R$?  It is not due to any symmetry and, unlike other theories,
cannot
be argued on the basis of renormalizability.  However physically the
curvature is small so that in most applications $R^2$ terms would be
yet
smaller.  This leads to the use of the energy expansion in the
gravitational effective field
theory.

There are in fact infinitely many terms allowed by general coordinate
invariance, i.e.,

\begin{equation}
S = \int d^4 x \sqrt{g} \left\{ \Lambda + {2 \over \kappa^2} R + c_1
R^2 + c_2
R_{\mu \nu} R^{\mu \nu} + \ldots + {\cal L}_{matter} \right\}
\end{equation}

\noindent Here the gravitational Lagrangians have been ordered in a
derivative expansion with $\Lambda$ being of order $\partial^0, R$ of
order $\partial^2, R^2$ and $R_{\mu \nu} R^{\mu \nu}$ of order
$\partial^4$ etc.  Note that in four dimensions we do not need to
include a
term $R_{\mu \nu \alpha \beta} R^{\mu \nu \alpha \beta}$ as the Gauss
Bonnet theorem allows this contribution to the action to be written in
terms
of $R^2$ and $R_{\mu \nu} R^{\mu \nu}$.

The first term in Eq.21 , i.e., $\Lambda$, is related to the
cosmological
constant, $\lambda = -8 \pi  G \Lambda$.
This is a term which in principle should be included, but
cosmology bounds $\mid \lambda \mid < 10^{-56} cm^{-2}, \mid \Lambda
\mid < 10^{-46} GeV^4$ so that this constant is unimportant at
ordinary
energies.  We then set $\Lambda = 0$ from now on.

In contrast, the $R^2$ terms are able to be shown to be unimportant in
a natural way.  Let us drop Lorentz indices in order to focus on the
important
elements, which are the numbers of derivatives.  A $R + R^2$
Lagrangian

\begin{equation}
{\cal L} = {2 \over \kappa^2} R + cR^2
\end{equation}

\noindent has an equation of motion which is of the form

\begin{equation}
\Box h + \kappa^2 c^2 \Box \Box h = 8 \pi GT
\end{equation}

\noindent The Greens function for this wave equation has the form

\begin{eqnarray}
G(x) & = & \int {d^4q \over (2 \pi)^4 } {e^{iq \cdot x} \over q^2
+ \kappa^2 c
q^4} \nonumber \\
& = & \int {d^4 q \over (2 \pi )^4} \left[ {1 \over q^2} - {1 \over
q^2 +
{1/\kappa^2 c}} \right] e^{-iq \cdot x}
\end{eqnarray}

\noindent The second term appears like a massive scalar, but with the
wrong
overall sign, and leads to a short-ranged Yukawa potential

\begin{equation}
V(r) = - G m_1 m_2 \left[ {1 \over r} - {e^{-r/ \sqrt{\kappa^2 c}}
\over
r}\right] .
\end{equation}

\noindent The exact form has been worked out by Stelle~\cite{stelle}, 
who gives
the
experimental bounds $c_1, c_2 < 10^{74}$.  Hence, if $c_i$ were a
reasonable number there would be no effect on any observable physics.
[Note that if $c \sim 1, \sqrt{\kappa^2 c} \sim 10^{-35}m]$.
Basically the
curvature is so small that $R^2$ terms are irrelevant at ordinary
scales.

As a slightly technical aside, in an effective field theory we should
not treat the $R^2$ terms to all orders,
as is done above in the exponential of the Yukawa solution, but only
include the
first corrections in $\kappa^2c$.  This is because at higher orders
in $\kappa^2c$ we
would also be sensitive to yet higher terms in the effective
Lagrangian
($R^3, R^4$ etc.) so that we really do not know the full $r
\rightarrow 0$
behavior.  Rather, for $\sqrt{\kappa^2c}$ small we can note the Yukawa
potential becomes a representation of a delta function

\begin{equation}
{e^{-r/ \sqrt{\kappa^2c}} \over r} \rightarrow 4\pi
\kappa^2c \delta^3 (\vec{r})
\end{equation}
The low energy potential then has the form
\begin{equation}
V(r) = -Gm_1 M_2 \left[{1 \over r} + 128 \pi^2 G (c_1 - c_2) \delta^3
(\vec{x}) \right]
\end{equation}

\noindent $R^2$ terms in the Lagrangian lead to a very weak and short
range modification to the gravitational interaction.

Thus when treated as a classical effective field theory, we can start
with the
more general Lagrangian, and find that only the effect of the Einstein
action,
$R$, is visible in any test of general relativity.  We need not make
any
unnatural restrictions on the Lagrangian to exclude $R^2$ and $R_{\mu
\nu} R^{\mu \nu}$ terms. 

\section{Quantization}

There is a beautiful and simple formalism for the quantization of
gravity.
The most attractive variant combines the covariant quantization
pioneered by
Feynman and De Witt~\cite{fdw} with the
background field method introduced in this
context by 't Hooft and Veltman~\cite{thv}.  The quantization of a gauge
theory
always involves fixing a gauge. This can in principle cause trouble if
this procedure then induces divergences which can not be absorbed
in the coefficients of the most general Lagrangian which displays the
gauge symmetry. The background field method solves this problem
because the calculation retains the symmetry under transformations of
the background field and therefor the loop expansion will be
gauge invariant, retaining the symmetries of general relativity.

Consider the expansion of the metric about a smooth background field
$\bar{g}_{\mu \nu} (x)$,

\begin{equation}
g_{\mu \nu} (x) = \bar{g}_{\mu \nu} (x) + \kappa h_{\mu \nu}
\end{equation}

\noindent Indices are now raised and lowered with $\bar{g}$.  The
Lagrangian may be expanded in the quantum field $h_{\mu \nu}$~\cite{thv}.

\begin{eqnarray}
{2 \over \kappa^2} \sqrt{g} R & = & \sqrt{\bar{g}} \left\{ {2
\over \kappa^2}
\bar{R} + {\cal L}_g^{(1)} + {\cal L}_g^{(2)} + \cdots
\right\}\nonumber
\\
{\cal L}^{(1)}_g & = & {h_{\mu \nu} \over \kappa} \left[ \bar{g}^{\mu
\nu}
\bar{R} - 2 \bar{R}^{\mu \nu} \right] \nonumber \\
{\cal L}^{(2)}_g & = & {1 \over 2} D_{\alpha} h_{\mu \nu} D^{\alpha}
h^{\mu \nu} - {1 \over 2} D_{\alpha} h D^{\alpha} h + D_{\alpha} h
D_{\beta} h^{\alpha \beta} \\
& & -D_{\alpha} h_{\mu \beta} D^{\beta} h^{\mu \alpha} + \bar{R}
\left(
{1 \over 2} h^2 - {1 \over 2} h_{\mu \nu}h^{\mu \nu} \right) \nonumber
\\
& & + \bar{R}^{\mu \nu} \left(2 h^{\lambda} _{~\mu} h_{\nu \alpha} - h
h_{\mu \nu} \right) \nonumber
\end{eqnarray}

\noindent Here $D_{\alpha}$ is a covariant derivative with respect to
the
background field.  The total set of terms linear in $h_{\mu \nu}$
(including
those from the matter Lagrangian) will vanish if $\bar{g}_{\mu \nu}$
satisfies Einstein's equation.  We are then left with a quadratic
Lagrangian
plus interaction terms of higher order.

However, the quadratic Lagrangian cannot be quantized without gauge
fixing and the associated Feynman-DeWitt-Fadeev-Popov
ghost fields.
In this case, we would like to impose the harmonic gauge constraint in
the
background field, and can choose the constraint~\cite{thv}

\begin{equation}
G^{\alpha} = \sqrt[4]{g} \left( D^{\nu} h_{\mu \nu} - {1 \over 2}
D_{\mu}
h^{\lambda} _{~\lambda} \right) t^{\nu \alpha}
\end{equation}

\noindent where

\begin{equation}
\eta_{\alpha \beta} t^{\mu \alpha} t^{\nu \beta} = \bar{g}^{\mu \nu}
\end{equation}

\noindent This leads to the gauge fixing Lagrangian~\cite{thv}

\begin{equation}
{\cal L}_{gf} = \sqrt{ \bar{g}} \left\{ \left( D^{\nu} h_{\mu \nu} -
{1 \over
2} D_{\mu} h^{\lambda} _{~\lambda} \right) \left( D_{\sigma} h^{\mu
\sigma} - {1 \over 2} D^{\mu} h^{\sigma} _{~\sigma} \right) \right\}
\end{equation}

\noindent Because the gauge constraint contains a free Lorentz index,
the ghost
field
will carry a Lorentz label, i.e., they will be fermionic vector
fields.
After a
bit of work the ghost Lagrangian is found to be

\begin{equation}
{\cal L}_{gh} = \sqrt{ \bar{g}} \eta^{* \mu} \left[ D_{\lambda}
D^{\lambda}
\bar{g}_{\mu \nu} - R_{\mu \nu} \right] \eta^{\nu}
\end{equation}

\noindent The full quantum action is then of the form

\begin{eqnarray}
S = \int d^4s \sqrt{\bar{g}} \left\{ {2 \over \kappa^2} \bar{R} - {1
\over 2}
h_{\alpha \beta} D^{\alpha \beta, \gamma \delta} h_{\gamma \delta}
\right.
\nonumber \\
+ \left. \eta^{* \mu} \left\{ D_{\lambda} D^{\lambda} \bar{g}_{\mu
\nu} - \bar{R}_{\mu \nu} \right\} \eta^{\nu} + {\cal O}(h^3) \right\}
\end{eqnarray}

\section{Renormalization}
The one loop divergences of gravity have been studied in two slightly
different methods.  One involves direct calculation of the Feynman
diagrams
with a particular choice of gauge and definition of the quantum
gravitational
field~\cite{direct}.  The background field method,
with a slightly different gauge constraint, allows one to calculate in
a single step
the divergences in graphs with arbitrary numbers of external lines and
also produces a result which is explicitly generally covariant~\cite{thv}.
In the latter technique one expands about a
background spacetime $\bar{g}_{\mu \nu}$, fixes the gauge as we
described above and collects all the terms quadratic in the quantum
field
$h_{\mu \nu}$ and the ghost fields.  For the graviton field we have

\begin{eqnarray}
Z \left[ \bar{g} \right] & = & \int \left[ dh_{\mu \nu} \right] exp
\left\{ i
\int d^4 x \sqrt{\bar{g}}  \left\{ {2 \over \kappa^2} \bar{R} + h_{\mu
\nu}
D^{\mu \nu \alpha \beta}  h_{\alpha \beta} \right\} \right. \nonumber
\\
& = & det D^{\mu \nu \alpha \beta} \nonumber \\
& = & exp Tr ln (D^{\mu \nu \alpha \beta})
\end{eqnarray}

\noindent where $D^{\mu \nu \alpha \beta}$ is a differential operator
made
up of derivatives as well as factors of the background curvature.  The
short
distance divergences of this object can be calculated by standard
techniques
once a regularization scheme is chosen.  Dimensional regularization is
the
preferred scheme because it does not interfere with the invariances of
general relativity.  First calculated in this scheme
by 't Hooft and Veltman~\cite{thv}, the divergent
term at one-loop due to graviton and ghost loops is described by a
Lagrangian

\begin{equation}
{\cal L}^{(div)}_{1 loop} = {1 \over 8 \pi^2 \epsilon} \left\{ {1
\over 120}
\bar{R}^2 + {7 \over 20} \bar{R}_{\mu \nu} \bar{R}^{\mu \nu} \right\}
\end{equation}

\noindent with $\epsilon = 4 - d$.  Matter fields of different spins
will also
provide additional contributions with different linear combinations of
$R^2$
and $R^{\mu \nu} R_{\mu \nu}$ at one loop.

The fact that the divergences is not proportional to the original
Einstein
action is an indication that the theory is of the non-renormalizable
type.
Despite the name, however, it is easy to renormalize  the theory at
any given
order.  At one loop we identify renormalized parameters

\begin{eqnarray}
c^{(r)}_1 & = & c_1 + {1 \over 960 \pi^2 \epsilon} \nonumber \\
c^{(r)}_2 & = & c_2 + {7 \over 160 \pi^2 \epsilon}
\end{eqnarray}

\noindent which will absorb the divergence due to graviton loops.
Alternate
but equivalent expressions would be used in the presence of matter
loops.

A few comments on this result are useful.  One often hears that pure
gravity
is one loop finite.  This is because the lowest order equation of
motion for
pure gravity is $R_{\mu \nu} = 0$ so that the ${\cal O} (R^2)$ terms
in the
Lagrangian vanish for all solutions to the Einstein equation.  However
in the
presence of matter (even classical matter) this is no longer true and
the
graviton loops yield divergent effects which must be renormalized as
described above.  At two loops, there is a divergence in pure
gravity which remains even after the equations of motion have been
used~\cite{twoloop}.

\begin{equation}
{\cal L}^{(div)}_{2 loop} = {209 \kappa^2 \over 2880 (16 \pi^2 )} {1
\over
\epsilon} \bar{R}^{\alpha \beta}_{~~\gamma \delta} \bar{R}^{\gamma
\delta}_{~~\eta \sigma} \bar{R}^{\eta \sigma}_{~~\alpha \beta}
\end{equation}

\noindent For our purposes, this latter result also serves to
illustrate the
nature of the loop expansion.  Higher order loops invariably involve
more
powers of $\kappa$ which by dimensional analysis implies more powers
of the
curvature or of derivatives in the corresponding Lagrangian (i.e., one
loop
implies $R^2$ terms, 2 loops imply $R^3$ etc.).  The two loop
divergence
would be renormalized by absorbing the effect into a renormalized
value of
a coupling constant in the ${\cal O}(R^3)$ Lagrangian.

\section{Quantum Predictions in An Effective Theory}

At this stage it is important to be clear about the nature of the
quantum
predictions in an effective theory.  The divergences described in the
last
section come out of loop diagrams, but they are {\em not} predictions
of
the effective theory.  They are due to the high energy portions of the
loop
integration, and we do not even pretend that this portion is reliable.
We
expect the real divergences (if any) to be different.  However the
divergences do not in any case enter into any physical consequences,
as
they absorbed into the renormalized parameters.  The couplings which
appear in the effective Lagrangian are also not predictions of the
effective
theory.  They parameterize our ignorance and must emerge from an
ultimate
high energy theory or be measured experimentally.  However there are
quantum effects which are due to low energy portion of the theory, and
which the effective theory can predict.  These come because the
effective
theory is using the correct degrees of freedom and the right vertices
at low
energy.  It is these low energy effects which are the quantum
predictions of
the effective field theory.

It may at first seem difficult to identify which components of a
calculation
correspond to low energy, but in practice it is straightforward.  The
effective field theory calculational technique automatically separates
the low
energy observables.  The local effective Lagrangian will generate
contributions to some set of processes, which will be parameterized by
a set
of coefficients.  If, in the calculation of the loop corrections, one
encounters
contributions which have the same form as those from the local
Lagrangian,
these cannot be distinguished from high energy effects.   In the
comparison
of different reactions, such effects play no role, since we do not
know
ahead of time the value of the coefficients in ${\cal L}$.  We must
measure these constants or form linear combinations of observables
which are independent of them. Only loop
contributions which have a different structure from the local
Lagrangian can make a difference in the
predictions of reactions.  Since the effective Lagrangian accounts for
the
most general high energy effects, anything with a different structure
must
come from low energy.

A particular class of low energy corrections stand out as the most
important.
These are the nonlocal effects.  In momentum space the nonlocality is
manifest by a nonanalytic behavior.  Nonanalytic terms are clearly
distinct
from the effects of the local Lagrangian, which always give results
which
involves powers of the momentum.

\section{Examples}

A conceptually simple (although calculationally difficult) example is
graviton-graviton scattering. This has been calculated to
one-loop in an impressive paper by Dunbar and Norridge~\cite{dn} 
using string
based methods. Because the reaction involves only the pure gravity
sector, and $R_{\mu\nu} = 0$ is the lowest order equation of motion,
the result is independent of any of the four-derivative terms that can
occur in the Lagrangian ($R^2$ or $R_{\mu\nu}R^{\mu\nu}$). Thus the
result is independent of any unknown coefficient to one loop order.
Their result for the scattering of positive helicity gravitons is
\begin{eqnarray}
{\cal{A}}(++\to ++) & = & 8\pi G {s^4 \over stu} \left\{ \right. 1
\nonumber  \\
 & + & {G \over \pi} \left[  \right. ~\left( t \ln ({-u\over \delta}) 
\ln ({-s
\over \delta}) + u \ln ({-t\over \delta }) \ln ({-s\over \delta}) 
+ s \ln ({-t \over \delta}) \ln ({-u\over \delta}) \right)
\nonumber \\
 & + & \ln ({t \over u}){tu(t-u)\over 60 s^6}\left( 341(t^4+u^4) +1609
(t^3u + u^3t) +2566 t^2u^2 \right)  \nonumber \\
 & + & \left( \ln^2 ({t\over u}) + \pi^2\right){tu(t+2u)(u+2t)\over
2s^7} \left(2t^4 + 2u^4 +2t^3u +2u^3t - t^2u^2\right)   \nonumber \\
 & + & {tu \over 360 s^5} \left( 1922(t^4 + u^4) +9143(t^3u+u^3t) 
+14622 t^2u^2 \right) \left. ~ \right]
\left. ~ \right\} \ \ 
\end{eqnarray}
\noindent where $s=(p_1+p_2)^2$, $t=(p_1-p_3)^2$, $u=(p_1-p_4)^2$,  
($s+t+u=0$)
and where I have used $\delta$ as an infrared cutoff~\cite{dt2}. 
One sees the
non-analytic terms in the logarithms. Also one sees the nature of the
energy expansion in the graviton sector - it is an expansion in $G
E^2$ where E is a typical energy in the problem. I consider this
result to be very beautiful. It is a low energy theorem of quantum
gravity. The graviton scattering amplitude {\it must} behave in this
specific fashion no matter what the ultimate high energy 
theory is and no matter
what the massive particles of the theory are. This is a rigorous
prediction of quantum gravity.

The other complete example of this style of calculation is the long
distance quantum
correction to the gravitational interaction of two 
masses~\cite{jfd2,others}. 
This is accomplished by calculating the vertex and vacuum polarization
corrections to the interaction of two heavy masses. In addition
to the classical
corrections~\cite{classical}, one obtains the true quantum correction 
\begin{equation}
V_{1pr}(r) = -{Gm_1m_2 \over r}\left[ 1 - {135 + 2 N_\nu \over
30 \pi^2 }{G \hbar \over r^2 c^3} +\ldots \right]
\end{equation} 
\noindent for a specific definition of the potential. Note that the
result is finite and independent of any parameters. This is easy to
understand once one appreciates the 
structure of effective field theory. 
The divergences
that occur in the loop diagrams all go into the renormalization of the
coefficients in the local Lagrangian, as we displayed above. Since
these terms in the Lagrangian yield only delta-function modifications
to the potential, they cannot modify any power-law correction that
survives to large distance. Only the
propagation of massless fields can generate
the nonanalytic behavior that
yields power-law corrections in
coordinate space. Since the low energy
couplings of massless particles are determined by Einstein's theory,
these effects are rigorously calculable.

Note that this calculation is the first to provide
a  quantitative answer to 
the question as to whether the effective gravitational coupling
increases or decreases at short distance due to quantum effects. While
there is some arbitrariness in what one defines to be $G_{eff}$, it
must be a universal property (this eliminates from consideration
the Post-Newtonian classical correction which depends on the external
masses) and must represent a general property of the theory. The
diagrams involved in the above potential are the same ones that go
into the definition of the running coupling in QED and QCD and the
quantum corrections are independent of the external masses. If one
uses this gravitational interaction to define a running coupling one
finds
\begin{equation}
G_{eff}(r) = G \left[ 1 - {135 + 2 N_\nu \over
30 \pi^2 }{G \hbar \over r^2 c^3}  \right]
\end{equation}
\noindent The quantum corrections {\it decrease} the strength of
gravity at short distance, in agreement with handwaving expectations.
(In pure gravity without photons or massless neutrinos, 
the factor $135 +2N_\nu$ is replaced by $127$.)
An alternate definition including the diagrams calculated 
in~\cite{others} has
a slightly different number, but the same qualitative conclusion. The
power-law running, instead of the usual logarithm, is a consequence of
the dimensionful gravitational coupling. 
 
These two results do not exhaust the predictions of the effective
field theory of gravity. In principle, any low energy gravitational
process can be calculated~\cite{hawking}. The two examples above have been
particularly nice in that they did not depend on any unknown
coefficients from the general Lagrangian. However it is not a failure
of the approach if one of these coefficients appears in a particular
set of amplitudes. One simply treats it as a coupling constant,
measuring it in one process (in principle) and using the result in the
remaining amplitudes. The leftover structures aside from this
coefficient are the low energy quantum predictions.

\section{Limitations and the high energy regime}
The effective field theory techniques can be applied at low energies
and small curvatures.
The techniques fail when the energy/curvature reaches the Planck
scale. There is no known method to extend such a theory to higher
energies. Indeed, even if such a technique were found, the result
would likely be wrong. In all known effective theories, new degrees of
freedom and new interactions come into play at high energies, and to
simply try to extend the low energy theory to all scales is the wrong
thing to do~\cite{comment}. One needs a new enlarged theory at high energy. However,
many attempts to quantize general relativity ignore this distinction
and appear misguided from our experience with other effective field
theories. While admittedly we cannot be completely sure of the high
energy fate of gravity, the structure of the theory itself hints very
strongly that new interactions are needed for a healthy high energy
theory. It is likely that, if one is concerned with only pure general
relativity, the effective field theory is the full quantum content of
the theory. 

\section{Summary}

The quantum theory of general relativity at low energy 
has turned out to be of the form that we call effective
field theories. 
The result is a beautiful theory that incorporates 
general coordinate invariance in a simple way, and 
which has a known methodology for extracting
predictions. The theory fits well with the other
ingredients of the Standard Model.  It is common, but wrong, 
to imply that general relativity differs for the other
interactions because it has no known quantum theory.
As we have seen, the quantum theory exists at those scales where
General Relativity is reliably thought to apply. 

Many of the most interesting questions that we ask of quantum 
gravity cannot be answered
by the effective field theory.  This is a warning that these questions
require knowledge of physics beyond the Planck scale. Since 
physics is an experimental science, thoughts about what 
goes on at such a high scale may remain merely speculation 
for many years.  However, it is at least reassuring that the 
ideas of quantum field theory can successfully be applied to 
General Relativity at the energy scales that we know about.


\begin{thebibliography}{99}
\bibitem{comment} In the question and answer period, Bryce De Witt
asked a question that at first 
appears only marginally related, but in fact
goes to the heart of the reasoning of effective field theory. He
imagined that it has been proven that pure QED is a perfectly complete and
consistent theory in every way satisfying to mathematical physics, and
asked if we should care about such a result. Effective field theory
would say that as physicists we would not care much about this
demonstration since we know that QED does not exist in isolation in
the world but is part of a different, larger theory. Even that larger
theory is likely not the full story. That these theories work at
present scales is the fascinating and important result.
\end{thebibliography}
\end{document}